\documentclass[aps,prl,twocolumn,showpacs,floatfix,superscriptaddress,notitlepage]{revtex4-2}
\usepackage[utf8]{inputenc}
\usepackage{soul}
\usepackage[left=2cm,right=2cm]{geometry}
\usepackage{physics}
\usepackage{amsmath}
\usepackage{amsfonts}
\usepackage{amssymb}
\usepackage{mathtools}
\usepackage{xcolor}
\usepackage{graphicx}
\usepackage[caption=false]{subfig}
\usepackage{dsfont}
\usepackage{amsthm}
\usepackage[hidelinks]{hyperref}
\usepackage{textcomp}
\usepackage{url}

\hypersetup{%
  colorlinks=true,
  linkcolor=blue,
  citecolor=blue,
  urlcolor=blue,
}

\begin{document}

\title{Geophysical survey based on Hybrid Gravimetry using Relative Measurements and an Atomic Gravimeter as an Absolute Reference}
%\title{Geophysical Survey using an Atomic Gravimeter as an Absolute Reference}

\author{Nathan Shettell}
\affiliation{Centre  for  Quantum  Technologies,  National  University  of  Singapore,  Singapore 117543, Singapore}

\author{Kai Sheng Lee}
%\email{} %incase we want someone to be corresponding author
\affiliation{Centre  for  Quantum  Technologies,  National  University  of  Singapore,  Singapore 117543, Singapore}

\author{Fong En Oon}
\affiliation{Centre  for  Quantum  Technologies,  National  University  of  Singapore,  Singapore 117543, Singapore}

\author{Elizaveta Maksimova}
\affiliation{Centre  for  Quantum  Technologies,  National  University  of  Singapore,  Singapore 117543, Singapore}

\author{Christoph Hufnagel}
\affiliation{Centre  for  Quantum  Technologies,  National  University  of  Singapore,  Singapore 117543, Singapore}

\author{Shengji Wei}
\affiliation{Earth Observatory of Singapore, Nanyang Technological University, Singapore, 639798, Singapore}
\affiliation{Asian School of the Environment, 639798, Nanyang Technological University, Singapore, Singapore}

\author{Rainer Dumke}
\affiliation{Centre  for  Quantum  Technologies,  National  University  of  Singapore,  Singapore 117543, Singapore}
\affiliation{School  of  Physical  and  Mathematical  Sciences,  Nanyang  Technological  University,  637371, Singapore}

%\date{\today}

\begin{abstract}
Gravimetry is a versatile metrological approach in geophysics to accurately map subterranean mass and density anomalies. There is a broad diversification regarding the working principle of gravimeters, wherein atomic gravimeters are one of the most technologically progressive class of gravimeters which can monitor gravity at an absolute scale with a high-repetition without exhibiting drift. Despite the apparent utility for geophysical surveys, atomic gravimeters are (currently) laboratory-bound devices due to the vexatious task of transportation. Here, we demonstrated the utility of an atomic gravimeter on-site during a gravity survey, where the issue of immobility was circumvented with a relative spring gravimeter. The atomic gravimeter served as a means to map the relative data from the spring gravimeter to an absolute measurement with an effective precision of 7.7{\textmu}Gal. Absolute measurements provide a robust and feasible method to define and control gravity data taken at different sites, or a later date, which is critical to analyze underground geological units, in particular when it is combined with other geophysical approaches.
\end{abstract}

\maketitle

\emph{Motivation - } Atomic gravimeters are a highly sophisticated class of sensors which rely on matter-wave interferometry to accurately infer gravity from the acceleration of a free-falling test mass \cite{kasevich1991atomic, peters1999measurement, peters2001high, de2008precision}. Absolute measurements on the order of {\textmu}Gal have been demonstrated by numerous research groups \cite{wu2014investigation, zhou2015micro, menoret2018gravity, oon2022compact}, moreover, their performance in terms of sensitivity, long-term stability and accuracy, either rival or outperform state-of-art traditional sensors \cite{gillot2014stability, farah2014underground}. These characteristics are very appealing for geophysical surveys, as the combination of {\textmu}Gal precision and long-term stability enables the investigation of a variety of geophysical processes, such as geothermal activity \cite{sugihara2008geothermal, sofyan2015first, nishijima2016repeat, portier2018hybrid, omollo2023analysis}, volcanology \cite{d2008new, greco2012combining, greco2022long}, glacier ablation \cite{timmen2012observing}, ground deformation \cite{zerbini2002multi, greco2022long}, and aquifer analysis \cite{pool2008utility, davis2008time}.

Recently, there has been substantial progress on transforming laboratory-based atomic gravimeters, to commercially viable field products \cite{timmen2012observing, bidel2013compact, wu2014investigation, menoret2018gravity, chen2020portable, narducci2022advances}; notably by optimising the mechanical layout resulting in more compact systems \cite{fu2019new, oon2022compact} and the use of more efficient vibration cancellation technologies \cite{oon2022vibration}. Despite these improvements, geophysical surveys are a major obstacle for atomic gravimeters due to their complexity in logistics and harsh environmental conditions. Furthermore, a meaningful gravity survey requires a large and dense grid of data points for an elaborate mapping of the measurement data. 
%For example, a lengthy cool-down period, required temperature stability, protection from the elements, et cetera.
Gravity surveys using exclusively an atomic gravimeter as gravity sensor are limited to a small number of data points \cite{timmen2012observing, menoret2018gravity}.

The vast majority of gravity surveys employ spring gravimeters \cite{Niebauer2015, van2017geophysics, stolz2021superconducting, crossley1999network}, as they are significantly more compact and mobile. They can attain similar levels of precision to atomic gravimeters~\cite{Niebauer2015}. However, spring gravimeters are relative instruments and
%\textcolor{red}{require zeroing to an absolute scale, which we now refer to as calibration in the text.}
are designed to have a reference point value manually calibrated. Unfortunately, due to temperature dependencies within the hardware, mechanical wear-and-tear, and aging, this reference point value drifts over time (bias-drift) \cite{okiwelu2011strategies, schilling2015accuracy}. To circumvent bias-drifts, the reference point values need to be re-calibrated regularly, consequently, it is challenging to assiduously characterize gravitational changes due to prolonged geophysical processes. Instead, relative gravimeters are best suited for the detection and imaging of subterranean features \cite{bielik2002neo, hokkanen2007effects}, hydrogeology \cite{hokkanen2006hydrogeological, hokkanen2007effects, tanaka2013hydrological}, monitoring signals from earthquakes \cite{niebauer2011monitoring} or volcanic eruptions \cite{carbone2006analysis, carbone2007data, pivetta2023hydrological}, as well as analyzing the accuracy of Ocean loading and tidal models \cite{boy2003comparison, mikolaj2013first}.

\begin{figure*}[t!]
    \centering
    \hspace*{\fill}%
    \subfloat[Spring Gravimeter Data Collection]{
    \includegraphics[height=6cm]{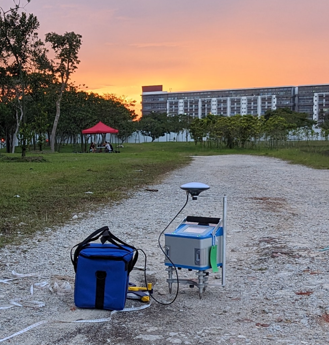}\label{fig:SpringGrav}
    }\hspace*{\fill}%
\subfloat[Atomic Gravimeter Housing]{
    \includegraphics[height=6cm]{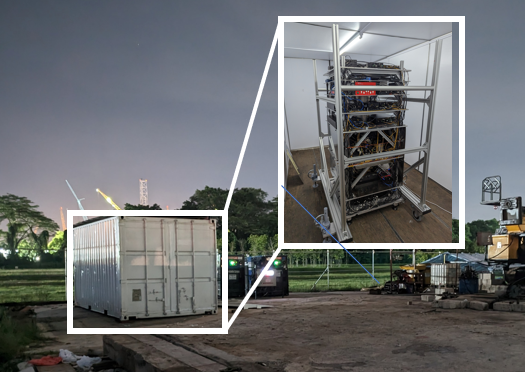}\label{fig:Housing}}\hspace*{\fill}%
\caption{Images from the gravity survey showcasing the (a) relative spring gravimeter (CG6), and the (b) absolute atomic gravimeter. (a) The CG6 measures approximately $30 \times 30 \times 30 \text{cm}^3$ and is mounted on a tripod with dials to accurately level the device. The spring gravimeter was equipped with a GPS antenna mounted by a custom made metal bracket fastened to the tripod. (b) The atomic gravimeter measures approximately $75 \times 75 \times 200 \text{cm}^3$ and was housed inside a repurposed shipping container. The shipping container was equipped with air conditioning in order to regulate the temperature within the operating conditions of the atomic gravimeter, as the outside temperature varied drastically from 25.0$^\circ$C to 34.4$^\circ$C, with a mean temperature of 28$^\circ$C, over the course of the two survey days.}
\label{fig:SurveyPictures}
\end{figure*}

By making use of an absolute gravimeter in tandem with a relative gravimeter, one can highlight the respective strengths of each technology while mitigating the aforementioned challenges concerning the limited mobility of the absolute atomic gravimeter, and the long-term stability of the spring gravimeter. This was the premise of a geophysical survey we conducted in Singapore, which took place over the course of two days. The relative gravimeter of choice was a quartz vertical spring gravimeter, dubbed the CG6, manufactured by \textit{Scintrex} \cite{hugill1990scintrex}. In quiet environments, the CG6 can attain a precision of 5-6{\textmu}Gal after an integration time of three minutes, which is sufficient to detect local deviations of gravity due to variations in subterranean features. Whereas the atomic gravimeter served as an absolute reference point, supplying a means to map from a relative value of gravity to an absolute value \cite{francis1998calibration, imanishi2002calibration, zerbini2002multi, sugihara2008geothermal, niebauer2011simultaneous, greco2012combining, lautier2014hybridizing, nishijima2016repeat}. Absolute measurements are necessary to properly analyze gravitational data taken at different sites or dates. For example, conducting large scale surveys over the period of a few days could suffer from ambiguity between bias-drift and gravitational changes due to environmental processes.

\begin{figure*}[t!]
    \centering
    \hspace*{\fill}%
    \subfloat[Atomic Gravimeter Measurements]{
    \includegraphics[width=0.48\textwidth]{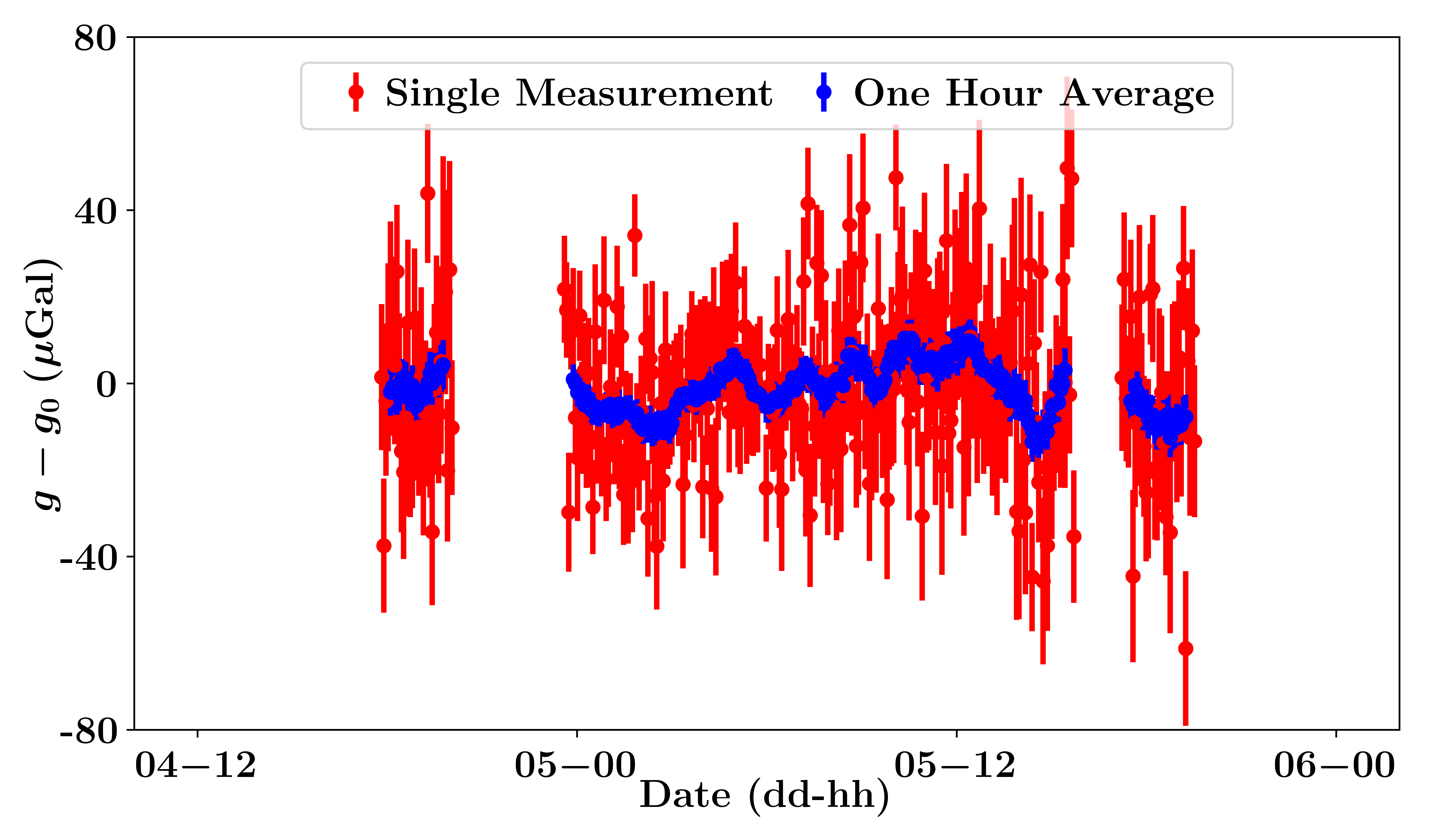}\label{fig:AtomMeas}
    }\hspace*{\fill}%
\subfloat[Allan Deviation]{
    \includegraphics[width=0.48\textwidth]{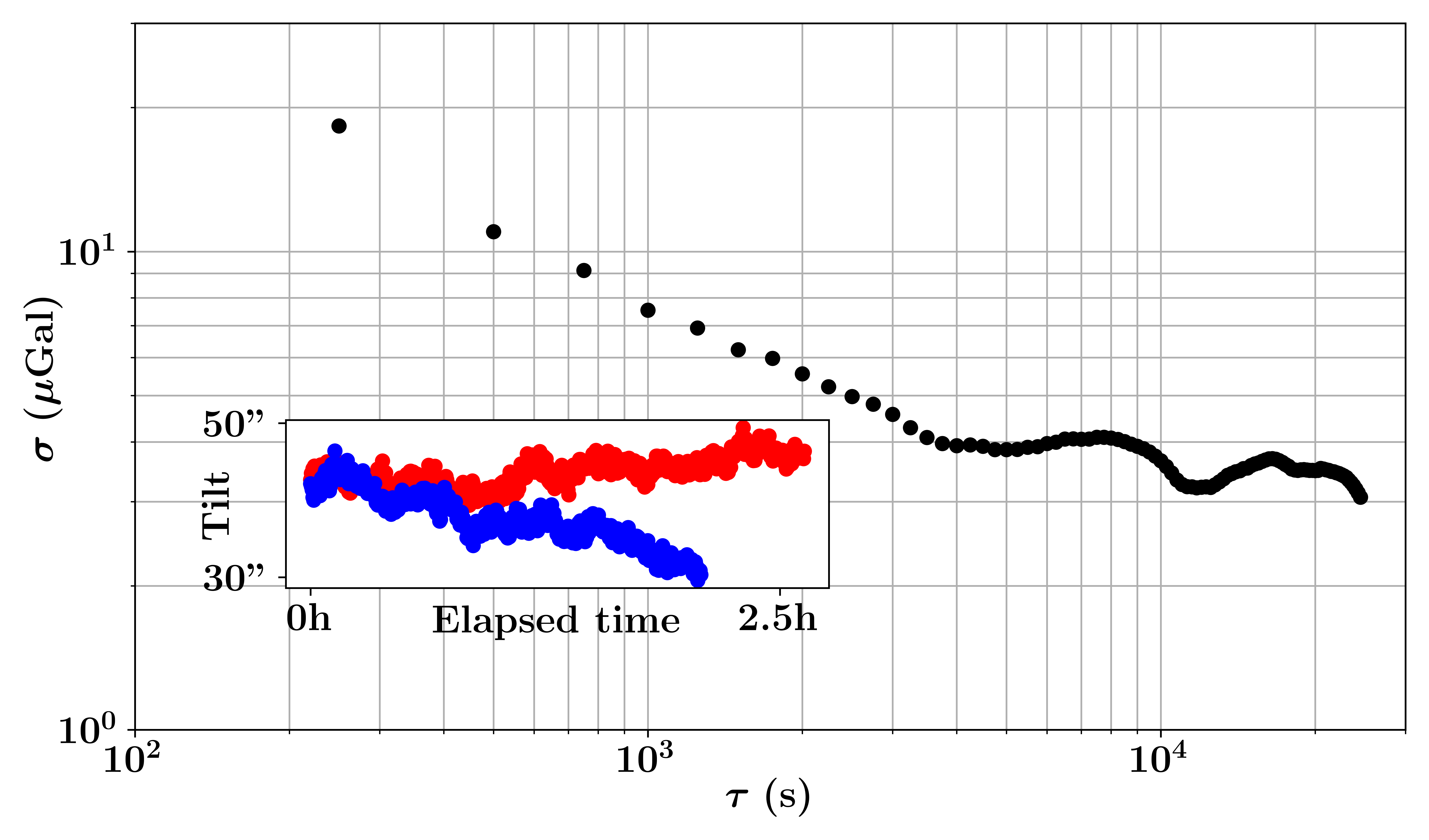}\label{fig:AllanVar}}\hspace*{\fill}%
\caption{(a) Results of the on-site atomic gravimeter measurements, in which changes due to tidal dynamics (computed using \textit{Quicktide Pro}) have been subtracted. The mean uncertainty of the single shot measurement (red) was $17.2\mu$Gal, which could be improved to $4.1\mu$Gal by taking one hour rolling average (blue). The data is shifted by $g_0=978061210${\textmu}Gal for visual clarity. (b) Allan deviation plot of the residual data, $g-g_\text{tide}$, in which the computation omitted any time-averages taken over periods with missing data. A comprehensive explanation of the Allan deviation calculation with dead times can be found in Ref.~\cite{sesia2008estimating}. The plateau in the stability profile of the gravimeter is due to the slow buckling of the floor below the atomic gravimeter. A tilt meter within proximity (whose readings are displayed on the bottom-left corner in arcseconds) measured a change in tilt of 6" on the first day of the survey (red) and 13" on the second day of the survey (blue).}
\label{fig:AtomGravResults}
\end{figure*}

\noindent\emph{Methods - } The survey was conducted over the course of two days, May 4th and May 5th 2023, at a geothermal exploration site in Singapore (longitude: 103.816933$^\circ$E, latitude: 1.458345$^\circ$N). Throughout the survey, local deviations of gravity were measured using a CG6. Each measurement spanned three minutes and, on average, attained a precision of 6{\textmu}Gal. The CG6 was equipped with a GPS antenna, which is portrayed in Fig.~(\ref{fig:SpringGrav}), whose position measurements were enhanced with real-time kinematic (RTK) corrections, allowing for the position of the gravity measurements to be made with an accuracy of 6mm with respect to longitude and latitude, and 8mm with respect to elevation. The physical spacing of the gravity measurements were approximately 25m apart, and followed a path parallel to a nearby road. The same path was repeated on both days to demonstrate the consistency of the measurements.

In addition to the gravity survey, a seismic survey was performed in parallel, in which the measurements had a similar placement profile. The methodology and results from the seismic survey can be found in Ref.~\cite{lythgoe2023fault}. The seismic survey was conducted by a different team, wherein the prospect of conducting two geophysical surveys was to compare two data sets of different origin to better characterize the geothermal activity being studied. Later in this manuscript a figure from the seismic survey is shown for comparative purpose.

Gravity discrepancies due to surface level deviations are filtered out by accounting for the varied elevation of the measurement sites. This is done by mapping the gravity measurements to an equipotential surface using the location of the base station as an elevation reference
\begin{equation}
    g \rightarrow g + g_{fa} - g_b,
\end{equation}
where $g_{fa} \approx 0.3086\text{mgal/m} \cdot \delta z$ is the free-air correction for a change in elevation of $\delta z$, and $g_b=2\pi \rho G \delta z$ is the Bouguer correction which corrects for the gravity signal emanating from surface-level terrain with density $\rho$. A value of $\rho = 1.599$g/cm$^3$ is used, which is the expected value for the local terrain composition. Note that the Bouguer correction is limited by an infinite-slab approximation \cite{chapin1996theory}. There exists more precise terrain corrections which deviates from the infinite-slab assumption, however, this is beyond the technical scope of this study.

%\textcolor{red}{In order to accurately map out gravitational anomalies, it is necessary to map the measurements onto a supposed equipotential surface which we have taken to be the sea-level. This was done by using the high-precision elevation measurements, $h$, to accurately correct for the elevation dependent Bouguer anomaly \cite{chapin1996theory}:
%\begin{equation}
%    \delta g = (\frac{2g}{R}-2 \pi G \rho) h,
%\end{equation}
%where the terms on the right refer to the free-air (due to the distance from the Earth's center of mass) and Bouguer corrections (due to the contribution of additional rocks) respectively. Here, we have used the values of 2g/R=0.3086 mGal/m and $\rho= $ kg/m$^3$ for the density of rock, determined from samples taken in the drilling project within the vicinity.}

During the survey, the atomic gravimeter was housed in a shipping container, as shown in Fig.~(\ref{fig:Housing}). The shipping container provided necessary temperature and humidity control, as well as protection from the elements (heavy rainfall was experienced on the afternoon of the 4th of May, 2023). A detailed description of our atomic gravimeter can be found in Refs.~\cite{oon2022compact,oon2022vibration}, and a more thorough explanation of the underlying physics is provided in Ref.~\cite{peters2001high}. In brief, the value of gravity is inferred from the interference pattern of a matter-wave interferometer. The measurement results are shown in Fig.~(\ref{fig:AtomMeas}), in which the precision of a single scan over an interference fringe (250s) is 17.2{\textmu}Gal, and 4.1{\textmu}Gal for a one-hour rolling average.

\begin{figure}[b]
    \centering
    \includegraphics[width=.48\textwidth]{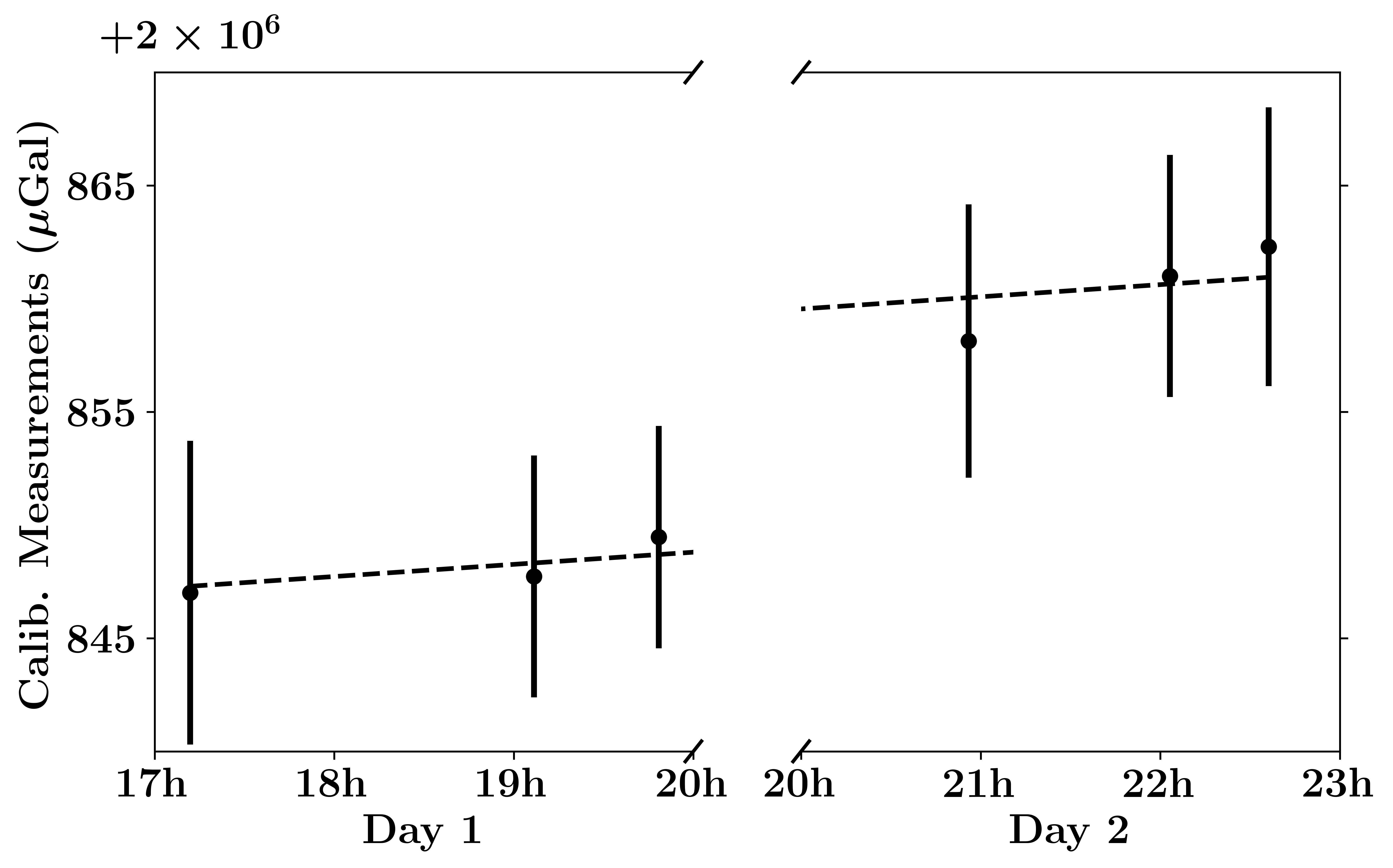}
    \caption{The recorded values of the CG6 during the calibration measurements. A line of best fit (dashed line) was used to determine that a correction of ($12.9 \pm 1.1$){\textmu}Gal/day should be applied to the field measurements (in addition to the pre-programmed drift correction) to better account for the inherent bias-drift of the machine.}
    \label{fig:calibration}
\end{figure}

A tilt meter was placed within proximity of the atomic gravimeter to monitor the level of the shipping containers floor. Unfortunately, the tilt meter observed a slow buckling in the flooring of the shipping container; a drift of approximately 6" on the first day of the survey, and 13" on the second day of the survey,  illustrated as a sub-plot within Fig.~(\ref{fig:AllanVar}). This drift translated to the atomic gravimeter readings, resulting a difference of -3.7{\textmu}Gal of the average measurements on the first and second day of the survey, hence the plateau exhibited on the Allan deviation curve. Additionally, the atomic gravimeter was subjected to significantly more noise than a typical laboratory environment, specifically due to an active drilling project within the vicinity (which is partially visible in Fig.~(\ref{fig:Housing})). These obstacles, slightly degrade one's ability to translate the measurements from the relative spring gravimeters to absolute values, however, only marginally when considering the precision of the relative gravity measurements. With additional precautions, the quality of an on-site absolute gravity measurements can be improved upon in future gravity surveys.

As a means to track the drift of the CG6 \cite{hugill1990scintrex}, six measurements were performed over the course of the survey (one at the start, middle, and end of each day). These measurements were taken at the same location, which was in proximity to the absolutely gravimeter. By determining the drift-rate of the CG6, the gravity measurements could be appropriately calibrated in post-processing, we henceforth refer to this set of measurements as `calibration measurements'. Although the CG6 is equipped with a built-in drift correction, this value is prone to errors due to higher order drift-rates and inaccuracies with the pre-programmed tidal correction \cite{okiwelu2011strategies} (which is elaborated upon in a later section of this manuscript). Additionally, the drift-rate may have incrementally changed due to transportation from the laboratory to the survey location \cite{mikolaj2013first, okiwelu2011strategies, iresha2021impact}. The readings taken during calibration tests are displayed in Fig.~(\ref{fig:calibration}), in which it was determined that the pre-programmed drift rate of CG6 was off by a rate of  ($12.9 \pm 1.1$){\textmu}Gal/day. In addition to drift correction, the calibration measurements provide a means of mapping the relative gravity measurements of the CG6 to an absolute value.

%Six calibration measurements were performed over the course of the survey (one at the start, middle, and end of each day). These measurements served dual purposes: to determine a mapping from the relative measurements of the gravimeter to an absolute value, as well as a means to correct for the linear drift inherent to the CG6 \cite{hugill1990scintrex}. Although the CG6 is equipped with a built-in drift correction, this value is prone to errors due to inaccuracies with the pre-programmed tidal correction \cite{okiwelu2011strategies} (which is elaborated upon in a later section of this manuscript). Additionally, the drift-rate may have incrementally changed due to transportation from the laboratory to the survey location \cite{mikolaj2013first, okiwelu2011strategies, iresha2021impact}. The readings taken during calibration tests are displayed in Fig.~(\ref{fig:calibration}), in which it was determined that the pre-programmed drift rate of CG6 was off by a rate of  ($12.9 \pm 1.1$){\textmu}Gal/day.

\begin{figure*}[t!]\centering
\subfloat[Gravity Survey Heat Map]{\label{fig:HeatMap}\includegraphics[height=6cm]{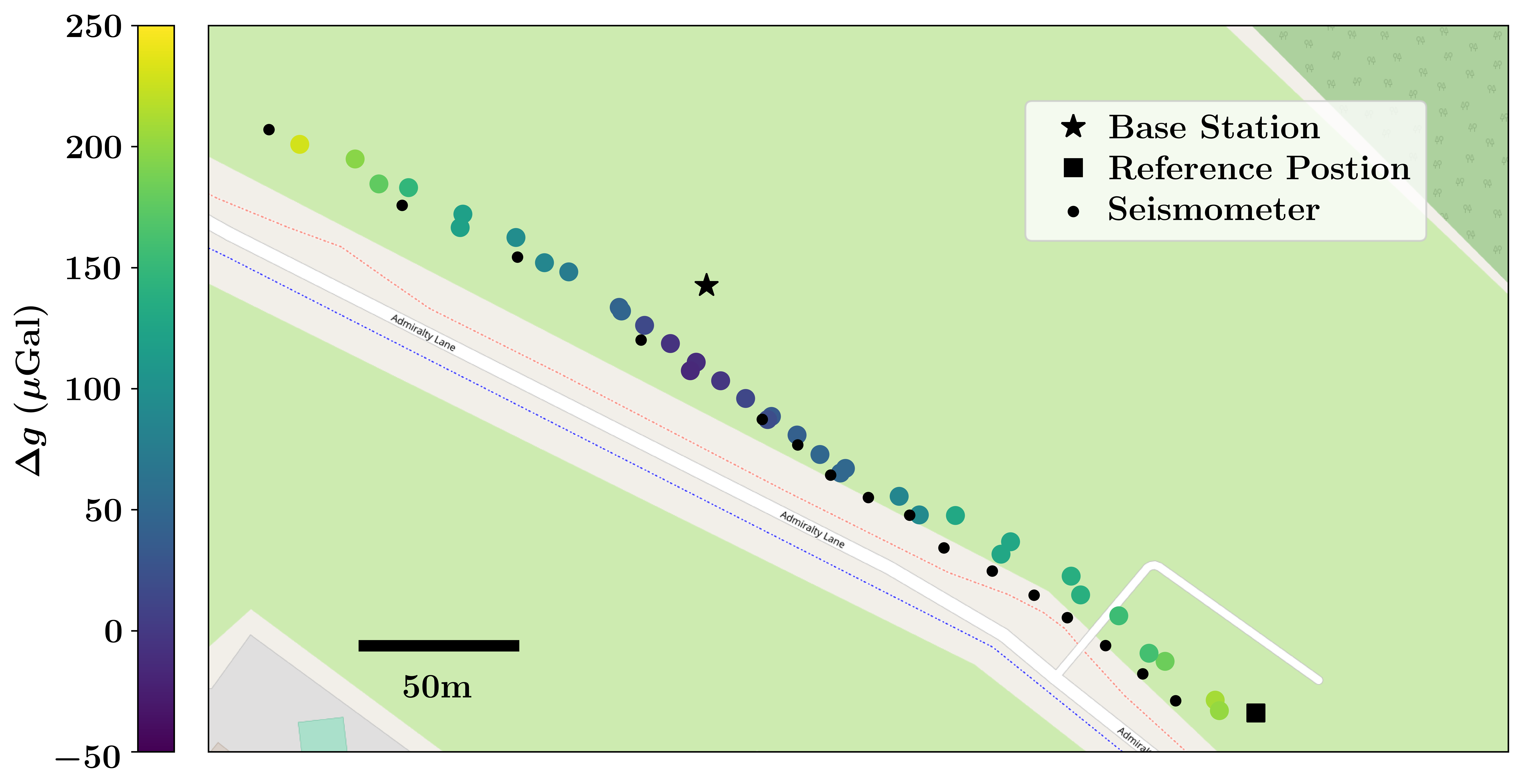}}\hfill\par
\hspace{0.08\textwidth}
\subfloat[Horizontal Gravity Gradient]{\label{fig:GravGrad}
\includegraphics[height=6cm]{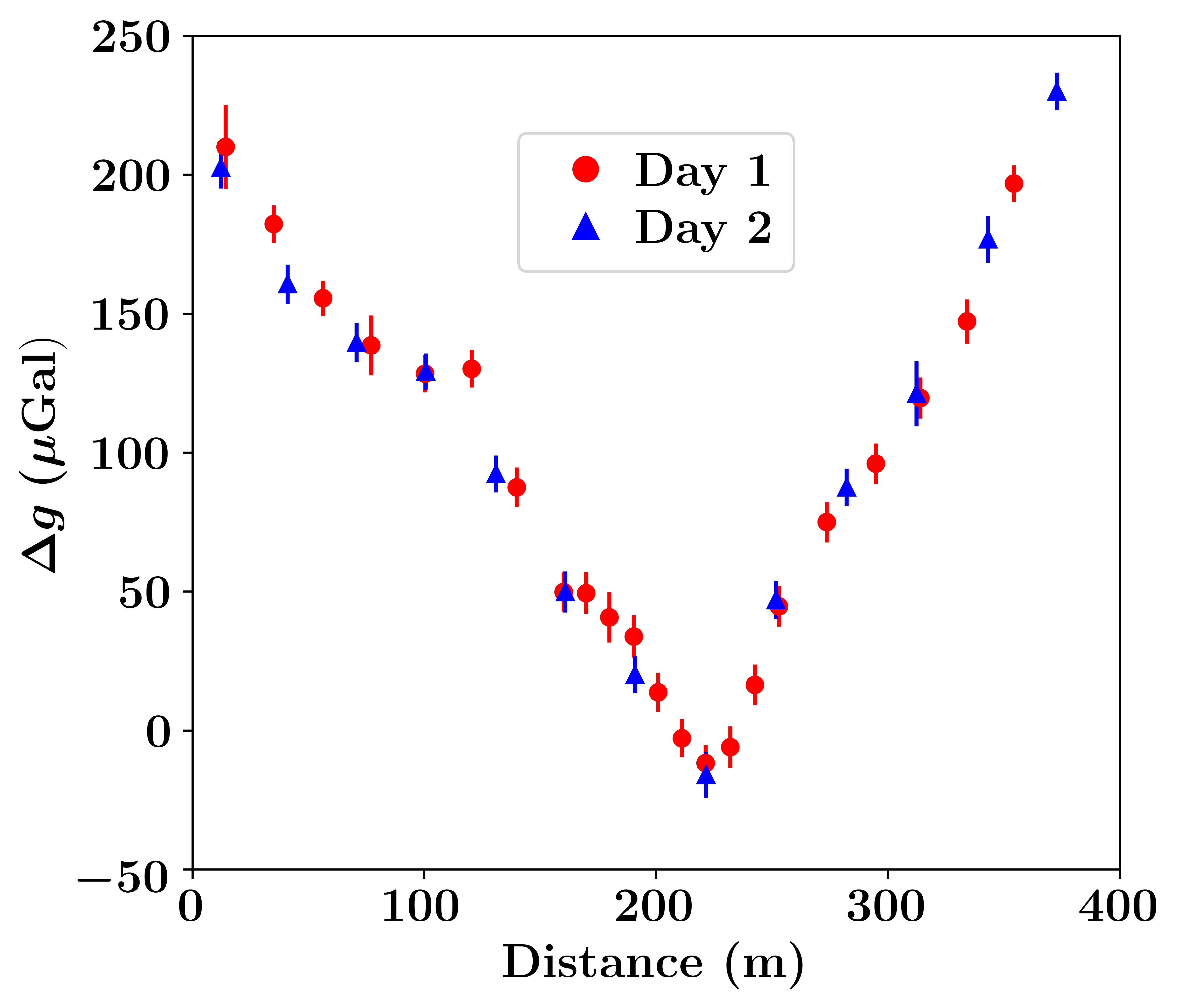}}\hfill 
\subfloat[Seismic Reflectivity]{\label{fig:SeismicData}\includegraphics[height=6cm]{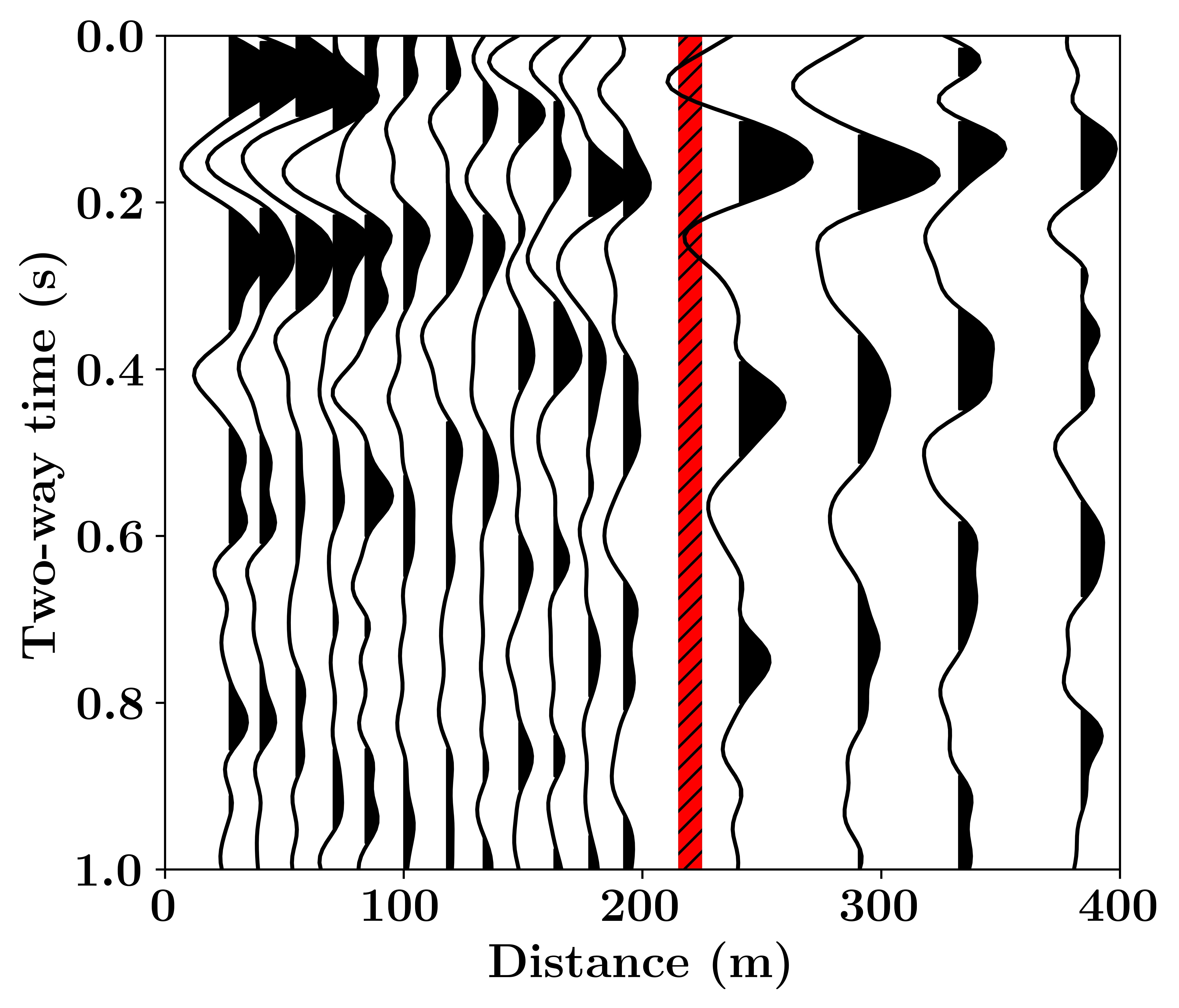}}\hspace{0.08\textwidth}
\caption{Results from the gravity geophysical survey at the NTU Geothermal site in Singapore. The values presented are relative to the mean relative value (with post-processing) at the base station. Corrections for tidal dynamics, free-air and Bouguer discrepancies \cite{chapin1996theory}, and additional drift determined by the calibration measurements (see Fig.~(\ref{fig:calibration})), have been applied. The gradiometry results are depicted via a heat map (a), as well as on a traditional Cartesian plot (b) with respect to the distance from the square marker depicted on the heat map. The results suggest a subterranean anomaly in proximity to the base station. This is further corroborated by the seismic reflectivity (c) observed by the seismology survey conducted in parallel \cite{lythgoe2023fault}, whose sensor placements are shown on the heat map. Here, the gravitational minimum is marked by a red bar.}
\label{fig:MainResults}
\end{figure*}

Combining the measurements taken by the absolute gravimeter, Fig.~(\ref{fig:AtomMeas}), along with the CG6 calibration measurements, Fig.~(\ref{fig:calibration}), we devise a formula to map relative gravity measurements at a position $\vec{r}$ and time $t$ to an absolute measurement
\begin{equation}
    \label{eq:absmap}
    g_\text{rel}(\vec{r},t) \rightarrow g_\text{abs}(\vec{r})=g_\text{abs}^\text{(ref)}+g_\text{rel}(\vec{r},t)-g_\text{rel}^\text{(ref)}(t-t_0).
\end{equation}
We assign $g_\text{abs}^\text{(ref)}$ to be the average absolute reading on the first day of the survey; we exclude the second day of readings due to more significant drift in the readings caused by the buckling of the floor, see Fig.~(\ref{fig:AllanVar}). On the other hand, $g_\text{rel}^\text{(ref)}$ is defined by the line of best fit through six calibration measurements, and is linearly dependent on the elapsed time since the initial calibration measurement: $t-t_0$. The overall uncertainty of the quantity $g_\text{abs}^\text{(ref)}-g_\text{rel}^\text{(ref)}(t-t_0)$ thus increase with $t-t_0$, however, the increase is marginal and is constrained within 4.0-4.2{\textmu}Gal throughout the survey. 

Notice that the assigned absolute gravity measurements, $g_\text{abs}(\vec{r})$, are deemed to independent of time (as tidal phenomena has been subtracted). Importantly though, this should be understood in the context of the timescale of this specific survey that took place May 4-5, 2023. The effective gravity at the survey location may vary on a much longer timescale, which necessitates a repeat survey in the future; as previously indicated, absolute measurements in the {\textmu}Gal are critical to compare data sets taken at much later dates. Nevertheless, the assumption that $g_\text{abs}(\vec{r})$ is time-independent for duration of the survey is an approximation, as atmospheric fluctuations (air pressure and precipitation) \cite{riccardi2007efficiency, delobbe2019exploring} and Ocean-tide loading \cite{okiwelu2011strategies, van1987displacements} will have a time-dependent effect on the gravity measurements. However, these phenomena, when combined, would account for changes in gravity ranging from 0.5-2.5{\textmu}Gal, which is inconsequential when considering the effective precision of the relative measurements after being mapped to an absolute measurement.

\noindent\emph{Results - } The results of our gravity survey are displayed in Fig.~(\ref{fig:GravGrad}), presented as a Bouguer gravity anomaly, indicates a sharp dip of approximately 250{\textmu}Gal in the middle of the gravity profile. This result suggests the presence of a negative subterranean density anomaly, which could be accredited to a fault zone structure \cite{oliver2011engineered, tjiawi2012natural, lythgoe2020large, lythgoe2021seismic, lythgoe2023fault}: a discontinuity in underground rock mass, which can act as a path way for fluid to travel from the underground geothermal reservoir to the hot spring at the surface. Geothermal reservoirs are a promising source of renewable energy for Singapore; supplying the means for an ecological and sustainable energy source \cite{stefansson2000renewability, oliver2011engineered}, which could potentially reduce the net carbon emissions of the country.
By conducting similar gravity surveys in the future, one can accurately monitor the activity of the geothermal reservoir, by virtue of the usage of an absolute gravimeter, Eq.~\eqref{eq:absmap}, \cite{sugihara2008geothermal, sofyan2015first, nishijima2016repeat, omollo2023analysis}, as well as to constrain larger scale structures with an expanded survey.

The location of the gravitational anomaly is highly consistent with the data derived from the seismic data~\cite{lythgoe2023fault}, Fig.~(\ref{fig:SeismicData}). The reflectivity plot captures the nature of the seismic waves (3-8Hz) measured by the seismometers over the span of five weeks, whose locations are marked on Fig.~(\ref{fig:HeatMap}). Specifically, the auto-correlation function of a single sensor is plotted vertically, centered at its respective distance from the reference position; the data is scaled horizontally for visual clarity. In brief, the reflectivity demonstrates a discontinuity in vertical velocity near the measured gravitational minimum (depicted as a red bar). As a future perspective, the combination of gravitational and seismic data may be processed via inversion algorithms \cite{yu2007constrained, sun2016joint}, enabling more robust data analysis due to the absolute character of the data set.

%The agreement between two independent geophysical survey outcomes suggest high robustness of both measurements and data analysis.

%An ongoing study by NTU and Tumcreate has detected temperatures of 60-90$^\circ$C at depths of 1.1km near the Sembawang hot spring, and project that temperatures may reach up to 200$^\circ$C at depths of 4-5km \cite{StraitsTimes1, StraitsTimes2}.

The recorded uncertainty of the raw measurement data (pre-corrections) on the CG6 ranges from 4.7-7.3{\textmu}Gal (with an exception of two measurements with recorded uncertainties of 9.9{\textmu}Gal and 14.5{\textmu}Gal). The principle reason for the variable uncertainty in Fig.~(\ref{fig:MainResults}) was the reliability of the RTK corrections within the GPS. When functioning optimally, the reported uncertainty in the elevation datum was 8mm, which translates to an added uncertainty of 1.9{\textmu}Gal in the gravitational datum post free-air and Bouguer corrections \cite{chapin1996theory}. However, the GPS signal occasionally dropped in quality, with reported accuracy as poor as 4cm, which results in an added uncertainty of 9.7{\textmu}Gal. The median effective uncertainty in the corrected data presented in Fig.~(\ref{fig:MainResults}) is 6.4{\textmu}Gal. Note that the above is not an exhaustive list of sources of uncertainty: fluctuations in topsoil density can result in sub-{\textmu}Gal sources of error \cite{lynch1983review}, and changes in atmospheric conditions (namely the heavy rain immediately prior to the commencement of the measurements on the first day of the survey \cite{SGWeather}) may have biased the measurements on the first day of the survey by 0.5-1.5{\textmu}Gal \cite{delobbe2019exploring}. Finally, the median uncertainty of the relative measurements when mapped to an absolute measurement, Eq.~\eqref{eq:absmap} is, $\sqrt{6.4^2+4.2^2}=$7.7{\textmu}Gal. This increases marginally to 7.8{\textmu}Gal if one includes an added uncertainty because of the aforementioned rain.

%When comparing the data sets between the first day and second day of the survey, there are six pairs of measurement locations which are within five meters of each other. Of these six pairs of points, four of them are within 4{\textmu}Gal of each other, which is within the resolution of the instruments, however   from these pair of measurements points, the measured gravity is, on average, 3.6{\textmu}Gal higher on the first day of the survey. The observed decrease in gravity from day-to-day could be regarded as statistical noise, however, we hypothesize that the heavy rainfall of 13.9mm shortly prior to when we started collecting data on the first day of survey \cite{SGWeather}, is the root cause of a subtle difference in gravitational data \cite{delobbe2019exploring}.

\noindent\emph{Future Perspectives - } As previously mentioned, the primary purpose of an absolute gravitational reference is to supply a means to compare relative data taken at different survey sites, or at later dates. To a greater extent though, an absolute reference serves as a means to filter out temporal shifts in gravity due to tidal dynamics, and enabling the isolation of gravitational shifts due to subterranean variations. Currently, tidal dynamics are typically filtered out through the means of predictive software, in fact, we use \textit{QuickTide Pro} as a benchmark for the long-term stability of our atomic gravimeter in Fig.~(\ref{fig:AllanVar}). However, due to the non-homogeneous nature of Earth's shape and it's non-rigidity, these programs are ultimately bounded in precision \cite{van2003efficiency,zurek2012constraining,liu2019high}.

\begin{figure}
    \centering
    \includegraphics[width=.48\textwidth]{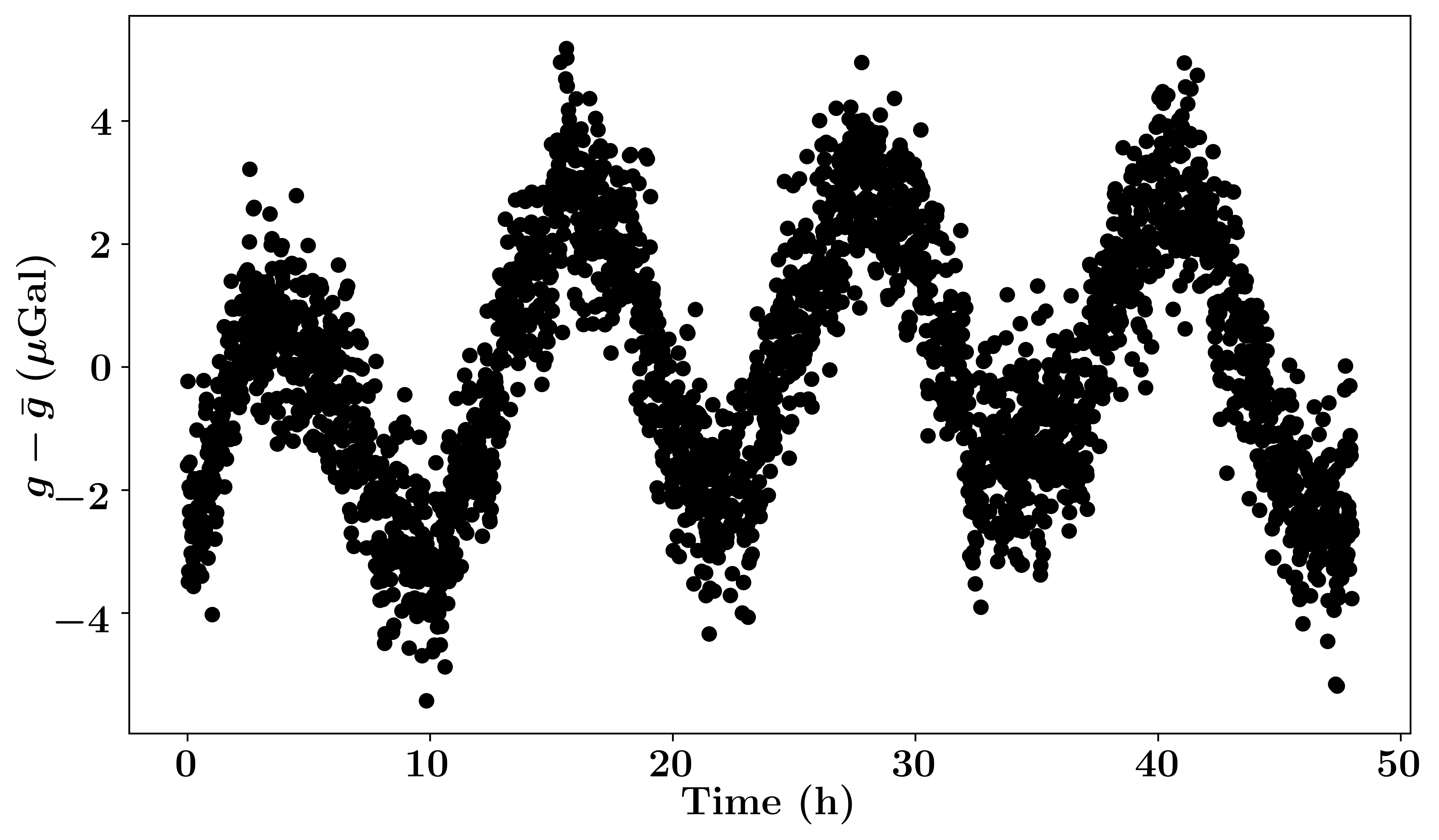}
    \caption{Two days of data collection with a stationary CG6 to demonstrate the errors when equipped with tidal correction software. The maximum difference from the mean value is $5.4${\textmu}Gal, and the standard deviation of the data set is $2${\textmu}Gal.}
    \label{fig:QTPEfficiency}
\end{figure}

%The inherent tidal correction of the CG6 is solely equipped with the Berger correction, thus, the gravitation signal observed can be used to analyze Ocean-tide loading in Singapore.

The accuracy of tidal correction software heavily depends on an underlying model \cite{van2003efficiency, jentzsch2005earth, matsumoto2001gotic2}, as well as geographic location. Notably, coastal locations are subject to more complex models, resulting in exacerbated errors when utilizing a slightly perturbed model \cite{okiwelu2011strategies, van1987displacements}. Being that Singapore is a diminutive island country, the pre-programmed tidal correction inherit to the CG6 demonstrates an error of up to 5.4{\textmu}Gal when taking measurements at Nanyang Technological Institute over the course of two days, see Fig.~(\ref{fig:QTPEfficiency}). As atomic gravimeters are capable of a {\textmu}Gal level of precision, they may be used in conjunction with tidal software for a reinforced tidal correction protocol.

%Tidal corrections are partitioned into two categories: i) the Berger correction which corrects for the relative positions of the Earth, Sun, and Moon \cite{van2003efficiency}, and ii) Ocean-tide loading which corrects for the non-rigidity of the Earth \cite{jentzsch2005earth, matsumoto2001gotic2}. The former correction is attributed to be accurate within 0.5{\textmu}Gal \cite{van2003efficiency}, whereas latter correction is significantly more complicated whose errors are exacerbated when in proximity the coast \cite{okiwelu2011strategies, van1987displacements}. Being that Singapore is a diminutive island country, Ocean-tide loading is omnipresent throughout Singapore; the gravitational magnitude of which was measured to be up to 5.4{\textmu}Gal at Nanyang Technological University - see Fig.~(\ref{fig:QTPEfficiency}). As atomic gravimeters continue to improve in precision, they may rival the precision of tidal forecasting models, enabling the utility of filtering out tidal behaviour from geophysical surveys.

\noindent\emph{Discussion - } Our gravity survey delineates the precision and effectiveness of utilizing an on-site atomic gravimeter alongside classical compact gravimeters during geophysical surveys. This hybridization is especially effective in monitoring slowly varying time-dependent signals, as observed in spatially and temporally resolved groundwater monitoring \cite{pool2008utility, davis2008time} or the progressive melting of polar ice caps \cite{timmen2012observing}. While classical gravimeters excel in survey applications, they tend to drift over time, a limitation ameliorated by the stability of atomic gravimeters, albeit at the cost of increased bulk which can be cumbersome for surveys.

The measurements from the geophysical survey indicate the existence of a subterranean anomaly beneath the NTU geothermal site in Singapore, which is hypothesized to be  a geothermal reservoir which resulted from the presence of a fault zone structure \cite{oliver2011engineered, tjiawi2012natural, lythgoe2020large, lythgoe2021seismic, lythgoe2023fault}. After performing necessary corrections, the median uncertainty in the relative data was 6.4{\textmu}Gal. Meanwhile, the atomic gravimeter collected data at a static position to provide a method to map the relative measurement values to an absolute value. After applying the map to the relative data set, the median uncertainty of the absolute measurements becomes 7.7{\textmu}Gal. This level of precision is necessary to accurately monitor the expansion of geothermal reservoirs with gravity measurements~\cite{sugihara2008geothermal, sofyan2015first, nishijima2016repeat, omollo2023analysis}.

In addition to geophysical and environmental applications, a {\textmu}Gal level of precision can be used to filter out tidal effects from gravity signals. Thus, it may be possible for next generation atomic gravimeters to present a pragmatic solution to curb unwanted tidal influences, instead of solely relying on software and mathematical models.
%a feat unattainable through predictive software constrained by underlying theoretical models.
This application is most useful in surveys in proximity to the coast, where tidal models are most prone to errors \cite{okiwelu2011strategies, van1987displacements}.

\bibliography{main}

\vspace{5pt}

\noindent\emph{Acknowledgements -} This research is supported by the NRF through NRF2021-QEP2-03-P06. The authors would like to thank Johnathan Kim for organizing the logistics of the geophysical survey, Karen Lythogoe for a fruitful discussion on seismology, as well as, Maung Maung Phyo, Yukuan Chen, Yu Jiang, and Nurdin Elon Dahlan for assisting with preparations of the geophysical survey.

\vspace{5pt}

\noindent\emph{Author Contributions -} N.S. and K.S.L. conducted the data analysis. K.S.L., F.E.O., E.M., C.H. and R.D. conducted the data collection during the survey. F.E.O. operated the atomic gravimeter. S.W. and R.D. conceived the project. N.S. and R.D. contributed to the writing of the manuscript.

\vspace{5pt}

\noindent\emph{Data Availability -} Raw data captured from the relative gravimeter, absolute gravimeter, and gps during the survey, can be downloaded from the following GitHub repository: github.com/GyroEmulator/SembawangRawData.

\vspace{5pt}

\noindent\emph{Competing Interests -} The authors have no conflicts to disclose.

%\clearpage
%\widetext
%\appendix 

\end{document}